\documentclass[a4paper,10pt]{article}

\def\eq#1{{Eq.~(\ref{#1})}}

\usepackage{latexsym}

\usepackage{graphicx,amssymb}
\usepackage{amsmath}
\newcommand{\cc}{cosmological constant}

\def\refcite{\cite}

\begin{document}

\markboth{T. Padmanabhan}{Gravity: A New Holographic Perspective}

%
%

\title{Gravity: A New Holographic Perspective}

\author{T. Padmanabhan\\
IUCAA, Pune University Campus,\\
Ganeshkhind, Pune 411007, INDIA\\
email: nabhan@iucaa.ernet.in}

\date{  }
\maketitle
\footnotetext[1]{Invited plenary talk delivered at the International Conference on {\it Einstein's Legacy in the New Millennium}, December 15 - 22, 2005, Puri, India.}

\begin{abstract}
A general paradigm for describing classical (and semiclassical) gravity is presented.
This approach brings to the centre-stage a holographic relationship between the 
bulk and surface terms in a general class of action functionals and 
provides a deeper insight into 
several aspects of classical gravity which have no explanation in the 
conventional approach. After highlighting a series of unresolved issues in
the conventional approach to gravity, I show that 
(i) principle of equivalence, 
(ii) general covariance and 
(iii) a reasonable condition on the variation of the 
  action functional, suggest  
    a generic Lagrangian for semiclassical gravity 
    of the form 
 $L=Q_a^{\phantom{a}bcd}R^a_{\phantom{a}bcd}$ with $\nabla_b\,Q_a^{\phantom{a}bcd}=0$. The 
 expansion of $Q_a^{\phantom{a}bcd}$ in terms of the derivatives of the metric tensor determines
 the structure of the theory uniquely. The zeroth order term gives the Einstein-Hilbert action and the first order correction
 is given by the  Gauss-Bonnet action. Any such Lagrangian can 
 be decomposed into a surface and bulk terms which are related holographically. The equations of motion
 can be obtained purely from a surface term in the gravity sector. Hence the field equations are invariant under the transformation $T_{ab} \to T_{ab} + \lambda g_{ab}$ and gravity does not respond to the changes in the bulk vacuum energy
  density. The cosmological constant arises as an integration constant in this approach. The implications are discussed.
  
\end{abstract}


\section{Why fix it when it works?}

Any attempt to provide a radically new perspective on gravity requires strong and clear motivation, since standard general relativity has been a very successful theory. So I will begin by providing the motivation for the alternative approach and identifying the ingredients it should have.

The elegance of general relativistic description of gravity rests on the geometric structure, which --- in turn --- is based on the Principle of Equivalence. I will interpret the Principle of Equivalence as allowing the description of gravity in terms of a metric tensor
and a compatible,  torsion-free, connection leading to the existence of local inertial frames around each event. This determines the kinematics of gravity (`how gravity tells matter to move') through the use of  special relativity in the local inertial frames. 
 
 Contrast this with the description of dynamics of gravity (`how matter tells spacetime to curve') for which we completely lack a similar guiding principle. There are several serious issues which crop up when we take a closer look at the issue of \textit{dynamics} of the gravitational field.
 
\subsection{Lack of foliation independent action functional} 
\label{sss:foliation}

Classical dynamics has to arise from semiclassical limit of quantum gravity through the 
extremisation of the  semiclassical action functional. \textit{Unfortunately, we  lack a guiding principle to choose such an action functional! } If $g_{ab}$ are the dynamical variables, the natural zeroth-order action functional should be quadratic in the derivatives $\partial g$ of $g_{ab}$. But Principle of Equivalence --- which allows us to reduce $g_{ab}\to\eta_{ab}, \partial_cg_{ab}\to0$ in any local region --- prevents the existence of such a generally covariant action. So, general covariance, combined with Principle of Equivalence, forces us to  include $\partial^2g$ terms in the action. \textit{This makes the variational principle  ill-defined and demands special treatment. }
For example, Einstein-Hilbert Lagrangian  $L_{EH}\propto R$ has a formal structure
\begin{equation}
 L_{EH}\sim R\sim (\partial g)^2+\partial^2g \equiv L_{\rm bulk} + L_{\rm sur} 
 \label{one}
\end{equation}
The surface term obtained by integrating $L_{sur}\propto \partial^2g$ should be ignored (or, more formally, canceled by an extrinsic curvature term; see e.g. \refcite{gh}) to obtain 
a well defined variational derivative that will lead to Einstein's equations.

So the (covariant) field equations  essentially arise from the variation of the non-covariant (or foliation dependent) bulk
term $L_{bulk}\propto (\partial g)^2$ --- usually called the $\Gamma^2$ Lagrangian. In classical theory, action principle is only a route to obtain the field equations and one might feel that --- as long as field equations have the required symmetries --- 
we need not worry whether the action principle has the correct symmetries. This, however, is a wrong attitude to adopt since  classical theory is just an approximation to the quantum theory and the form of the action functional ``off-shell" is important in quantum theory. Hence, one would have liked the action functionals --- and not just field equations --- to have the required symmetries.

This situation is unparalleled in any other theory in physics like e.g in the case of Yang-Mills field; the most natural action functional 
that one would write down for  gauge theories --- integral of the \textit{square} of the curvature, $F_{ab}F^{ab}$ --- is both gauge invariant and quadratic in the first derivatives of the dynamical variables (which are the connections $A_i$). In contrast, conventional gravity 
uses metric (rather than connection) as the basic dynamical variable and the action is \textit{linear} in the curvature [and nonpolynomial in the metric due to $\sqrt{-g}$ factor] 
thereby demanding a rather peculiar treatment of the action functional.

\subsection{Where are the degrees of freedom of gravity located?} 
\label{sss:dof}

 It is quite possible that continuum spacetime is like an elastic solid  and using $g_{ab}$ as a dynamical variable in quantum gravity is like using the density $\rho$ of of a solid as a fundamental variable in trying to obtain the quantum theory of solids. One might suspect that some of the difficulties in quantising gravity could be due to quantising the wrong action functional based on wrong fundamental variables. Hence identifying the correct action functional based on correct dynamical variables is important.

In using $L_{bulk}$ of \eq{one} to obtain the dynamics, we are also  assuming tacitly that the degrees of freedom are the components of the metric and they
reside in the spacetime volume $\mathcal{V}$. Such a description
in terms of $g_{ab}$ may be most geometrical but it is highly gauge redundant. Recall that,
 around any event, one can choose a local inertial frame so that $L_{bulk}\propto (\partial g)^2$ vanishes since $\partial g$ vanishes. On the other hand, one \textit{cannot} make $L_{sur}\propto\partial^2 g$ part to vanish by any choice of coordinates. 
 (This is most easily seen by evaluating $L_{\rm EH}$  in Riemann  normal coordinates in which
the bulk vanishes and only $L_{sur}$ contributes to $L_{EH}$).
 This suggests that \cite{tpholo} the true  degrees of freedom of gravity 
for a volume $\mathcal{V}$, which cannot be eliminated by a gauge choice,
reside in its boundary $\partial\mathcal{V}$ and contributes only to $L_{sur}$ around any event.

\subsection{Horizons and the gravitational degrees of freedom}
\label{sss:horizon}

The existence of horizons is generic in any geometric theory of gravity in which light rays are affected by the gravitational field.
In \textit{any} spacetime, there will exist families of observers (congruence of timelike curves) 
who will have access to only part of the spacetime. 
The boundary of the union of causal pasts of  the observers in a given congruence --- which is essentially  the boundary of the union of backward
light cones ---  will define a {\it causal} horizon $\mathcal{H}$ for this congruence. 
The well known examples are observers at $r=$ constant$>2M$
   in the Schwarzschild spacetime or the uniformly accelerated observers in flat spacetime.
This causal  horizon is
  \textit{dependent} on the family of observers that is chosen, but is \textit{coordinate independent}.
  
  Any class of observers, of course, has an equal right to describe physical phenomena \textit{entirely in terms of the variables defined in the regions accessible to them}. 
The action
functional describing gravity, used by these observers (who have access to only part of the spacetime)  can only depend on variables defined on the region accessible to them, including the boundary of this region. Since the horizon (and associated boundaries) may exist for some observers (e.g., uniformly accelerated observers in flat spacetime, $r=$ constant $>2M$ observers in the Schwarzschild spacetime ...) but not for others (e.g, inertial observers in flat spacetime, freely falling observers
inside the event horizon, $r<2M$, in the Schwarzschild spacetime ... ), this introduces a new level of observer dependence in the action functional describing the theory. I stress that this view point \textit{is completely in concordance with what we do in other branches of physics, while defining action functionals}. The action describing QED at $10$ MeV, say,
does not use degrees of freedom relevant at $10^{19}$ GeV which we have no access to. Similarly, if an observer has no access to part of the spacetime, (s)he should be able to use an action principle
using the variables (s)he can access. We are merely translating a well known paradigm
in momentum space to coordinate space. (Instead of integrating out the UV modes, we now need to integrate out the IR modes.)

The physics of the region blocked by the horizon has to be encoded in a boundary term in the action and we again see the importance of surface degrees of freedom and surface terms in action.

\subsection{Gravity-Thermodynamics connection}
\label{sss:tds}

The role of surface terms in the action functional, in determining the true degrees of freedom, is also strongly supported by the study
of horizon entropy, which shows that the degrees of freedom hidden by a horizon scales as the area and not as the volume.  Since any generic spacelike two surface can act as a horizon for \textit{some} class of observers, this again suggests that the gravitational degrees of freedom reside in the surface. (For a review of thermodynamics of horizons from this perspective, see  
ref. \refcite{tpreview}).

In the case of spherically symmetric spacetimes with $g_{00}=1/g_{rr}=-f(r)$
this leads to a deep (unexplained) connection between Einstein's equations and horizon thermodynamics \cite{ss}. Suppose that there is a horizon at $r=a$ so that $f(a)=0$ with the horizon temperature $k_BT=\hbar c f'(a)/4\pi$
determined by, say, Euclidean continuation. (We have reintroduced normal units temporarily). The Einstein's equation for this metric,
$(1-f)-rf'(r)=-(8\pi G/c^4) P$ (where $P$ is the radial pressure), evaluated at $r=a$ gives
\begin{equation}
\frac{c^4}{G}\left[{1\over 2} f'(a)a - \frac{1}{2}\right] = 4\pi P a^2
\label{reqa}
\end{equation} 
If we now consider two solutions with two different radii $a$ and $a+da$ for the horizon,
then multiplying the \eq{reqa} by $da$, and introducing a $\hbar$ factor \textit{by hand} into an otherwise classical equation, we can rewrite it as
\begin{equation}
   \underbrace{\frac{{{\hbar}} cf'(a)}{4\pi}}_{\displaystyle{k_BT}}
    \ \underbrace{\frac{c^3}{G{{\hbar}}}d\left( {1\over 4} 4\pi a^2 \right)}_{
    \displaystyle{dS}} 
  \ \underbrace{-\ {1\over 2}\frac{c^4 da}{G}}_{
    \displaystyle{-dE}}
 = \underbrace{P d \left( {4\pi \over 3}  a^3 \right)  }_{
    \displaystyle{P\, dV}}
\label{thermo}
\end{equation}
and read off the expressions:
\begin{equation}
 S={1\over 4L_P^2} (4\pi a^2) = {1\over 4} {A_H\over L_P^2}; \quad E={c^4\over 2G} a 
    =\frac{c^4}{G}\left( {A_H\over 16 \pi}\right)^{1/2}
\end{equation} 
where $A_H$ is the horizon area and $L_P^2=G\hbar/c^3$. Thus Einstein's equations can be cast as a thermodynamic identity. Two comments are relevant regarding this result:

(a) The combination $TdS$ is completely classical and is independent of $\hbar$ but $T\propto \hbar$
and $S\propto 1/\hbar$. This is analogous to the situation in classical thermodynamics when compared to statistical mechanics. The $TdS$ in thermodynamics is independent of Boltzmann's constant while statistical mechanics will lead to an $S\propto k_B$
and $T\propto1/k_B$. 

(b) In spite of superficial similarity, \eq{thermo} is different from the conventional first law of black hole thermodynamics, (as well as some previous attempts to relate thermodynamics and gravity, like e.g. the second paper in ref. \refcite{sakharov}), due to the presence of $PdV$ term. This relation  is more in tune with the membrane paradigm \cite{membrane} for the blackholes. This is easily seen, for example,  in the case of Reissner-Nordstrom blackhole
for which $P\neq0$. If a chargeless particle of mass $dM$ is dropped into a Reissner-Nordstrom blackhole, then an elementary calculation shows that the energy defined above as $E\equiv a/2$ changes by $dE= (da/2) =(1/2)[a/(a-M)]dM\neq dM$ while it is $dE+PdV$ which is precisely equal to $dM$
making sure $TdS=dM$. So we need the $PdV$ term to get $TdS=dM$ when a \textit{chargeless} particle is dropped into a Reissner-Nordstrom blackhole. More generally,
if $da$ arises due to changes $dM$ and $dQ$, it is easy to show  that \eq{thermo} gives $TdS=dM -(Q/a)dQ$ where the second term arises from the electrostatic contribution from the horizon surface charge as expected in the membrane paradigm.

Dynamically, \eq{thermo} is best interpreted as the energy balance under infinitesimal virtual displacements of the horizon normal to itself. This again brings up  the issue of surface degrees of freedom and will play a crucial role in our ensuing discussions.

\subsection{Gravity's immunity from vacuum energy}
\label{sss:gravimmu}

The clearest pointer against the conventional approach to gravity, based on the equation $G_{ab}=8\pi T_{ab}$,
is that we do not see gravitational effects due to vacuum energy density \textit{shifts} of the form
$T_{ab}\to T_{ab}+\Lambda g_{ab}$. Matter sector is invariant under shifting $L_{matter}$ by a constant but gravity sector is not! Every phase transition in the early universe produces such a change but it does not lead to gravitational effects.

The only way out of this problem --- which is logically distinct from standard cosmological constant problem, though related --- 
is to change field equations \cite{cc1} such that they are invariant under $T_{ab}\to T_{ab}+\Lambda g_{ab}$. This is same as working with \cite{unimod} the trace-free part $R_{ab}-(1/4)Rg_{ab}=8\pi(T_{ab}-(1/4)Tg_{ab})$ or, equivalently, with the equations
$(G_{ab}-8\pi T_{ab})\xi^a\xi^b=0$ for all \textit{null} vectors $\xi^a$. Either formulation, when combined with Bianchi identity, leads to $G_{ab}=8\pi T_{ab} +\lambda g_{ab}$ with some undetermined integration constant $\lambda$. The $\lambda$  is no longer a term in the equations but is part of the solution --- like $M$ in the Schwarzschild metric. One is free to choose it differently in different contexts, depending on physical situation. 

While this does not ``solve" the cosmological constant problem, it changes its nature completely because the theory is now invariant under $T_{ab}\to T_{ab}+\Lambda g_{ab}$. 
 (For a review of the cosmological constant problem from different perspectives, see ref.  \refcite{ccreviews}.)
Obviously, the conventional action principle with the gravitational degrees of freedom residing in the bulk cannot give raise to this but if we have only surface degrees of freedom, it seems plausible that the gravity will be unaffected by bulk vacuum energy. In fact, the 
shift from volume degrees of freedom to area degrees of freedom  can change \cite{cc1,cc2}
  the effective 
  energy density of the vacuum that is coupled to gravity from the gigantic $L_P^{-4}$ to the observed value
 $L_P^{-4} (L_P^2/H^{-2})$  and can lead to the observed value of the cosmological constant. 
 
\subsection{Holography of gravitational action}
\label{sss:holo}

Each of the features described above suggests a description of gravity based on surface degrees of freedom. Since the discussion in Sec.\ref{sss:gravimmu} suggests that such an approach should lead to the same equations of motion (except for a cosmological constant) as derived from the $L_{bulk}$,
one suspects that the surface and bulk terms of a gravitational action must encode the same amount of dynamical content. This suspicion is strengthened by the following remarkable relation  \cite{tpholo,tpreview} between the two parts of the Lagrangian in the Einstein-Hilbert action: 
\begin{equation}
\sqrt{-g}L_{sur}=-\partial_a\left(g_{ik}\frac{\partial \sqrt{-g}L_{bulk}}{\partial(\partial_a g_{ik})}      \right)
\label{surbulkrel}
\end{equation} 
Because of the existence of this relation, the transition from $L_{bulk}$ to $L_{EH}=L_{bulk}+L_{sur}$ can  be thought of as a transition from co-ordinate representation to momentum representation.  Given any \(L_q(\dot q,q)\), we can always construct a \(L_p(\ddot q,\dot q, q)\) which \textit{depends on the second derivatives} $\ddot q$ --- but gives the same equation of motion ---
  by using
\begin{equation}
 L_p=L_q-\frac{d}{dt}\left(q\frac{\partial L_q}{\partial\dot q}\right)
\end{equation} 
   Keeping     {{\(\delta p = 0\)}} at the end points and   varying \(L_p\)
 leads to the the same equations of motion  as  keeping  {{\(\delta q =0\)}} at the end points and
 varying \(L_q\).  
 In quantum theory, the path integral with \( L_p\)  gives the momentum space kernel \(G(p_2,t_2;p_1,t_1)\) just as path integral
 with \( L_q\)  gives the co-ordinate space kernel \(K(q_2,t_2;q_1,t_1)\).
Relation of this kind clearly indicates that both $L_{sur}$ and $L_{bulk}$ contain the same information content. In the conventional approach to gravity, \eq{surbulkrel} has no simple explanation; it is not clear why the surface and bulk terms should be related by  \eq{surbulkrel}
if the total action has to be generally covariant.

\subsection{Lack of a guiding principle to determine semiclassical dynamics }
\label{sss:guide}

Finally, note that we have no clear guiding principle to determine the action functional for gravity, something which we mentioned in Sec.\ref{sss:foliation}. The situation gets worse when we realise that the description in terms of metric is only a low-energy effective description.  The semiclassical theory is likely to exist in some $D$ dimensional spacetime with $D>4$ and quantum corrections will add  higher order correction terms involving squares of the curvature etc. We desperately need a guiding principle or symmetry to determine these higher order terms.

So the answer to ``why fix it while it works?" is that, if you listen carefully, you hear  creaking noises from the machinery all over the place.

\section{Gravity from the surface degrees of freedom}

I will now describe a paradigm in which all these issues can be successfully tackled. After briefly discussing how standard Einstein's theory arises from an  action functional  containing \textit{only} a surface term \cite{newper}, I will develop a general frame work for the low energy gravitational action functional which obeys principle of equivalence and general covariance.
I will show that, under very general conditions, such actions  have a generic
 structure, using which one can systematically obtain semi classical corrections to classical gravity. When the action is  expanded in the powers of
  curvature, one obtains  the Einstein-Hilbert (EH) action as the \textit{unique} zero-order term along with Gauss-Bonnet
  (GB) type correction as the \textit{unique} first-order term \cite{cc1,gr06}. What is probably even more remarkable is that \textit{all these action functionals allow a natural,
  holographically dual, description}; that is, the action functional can be expressed as a sum of bulk and 
  surface terms and the field equations can be obtained either from the surface term or from the bulk term. The gauge redundancy of 
  geometric description therefore allows all these theories to be described \textit{entirely} in terms
  of surface degrees of freedom thereby providing --- among other things --- a new backdrop to view the cosmological
  constant problem.  
 
 I begin with the result (see ref. \refcite{newper} for details) that it is possible to obtain the Einstein equations from an approach which  uses \textit{only} a surface term; \textit{ we do not need the bulk term at all!}.
In this approach, the action functional  is
\begin{equation}
A_{tot}=A_{sur}+A_{matter}=
\frac{1}{16\pi G}\int_{\partial\mathcal{V}} d^3 x \, 
\sqrt{-g}n_cQ_a^{\phantom{a}bcd}\Gamma^a_{bd}
+\int_{\mathcal{V}} d^4x \, \sqrt{-g}L_{m}(g,\phi)
\label{actfunc}
\end{equation}
where $Q_a{}^{bcd}=(1/2)(-\delta^c_ag^{bd}+\delta^d_ag^{bc})$.
Matter degrees of freedom live in the bulk $\mathcal{V}$ while the gravity contributes on the boundary ${\partial\mathcal{V}}$. (The surface term is the same as the surface term $L_{sur}$ in Einstein-Hilbert action except for a sign-flip.)
We demand that, whenever the boundary ${\partial\mathcal{V}}$
has a part $\mathcal{H}$ which is a null surface 
 the action should be invariant under $\delta g^{ab}=\nabla^a\xi^b+\nabla^b\xi^a$, where
 $\xi^a$ is the null normal to  $\mathcal{H}$ and is nonzero only on $\mathcal{H}$. \footnote{As far as gravity is concerned, this can be thought of as arising due to 
 virtual displacements of this horizon normal
 to itself. But note that in the matter lagrangian we are only varying $\delta g^{ab}=\nabla^a\xi^b+\nabla^b\xi^a$ and \textit{not} the matter fields $\phi$. In the case of a genuine active coordinate transformation, even matter fields will change, which is \textit{not} the case we are considering. We merely demand $\delta A_{tot}=0$ for a particular type of variation $\delta g_{ab}$ keeping everything else fixed.}
 This leads to Einstein's theory \textit{with a cosmological constant appearing as an integration constant} \cite{cc1,newper}. Since this should hold at every null surface through every event, the field equations hold at every event.

Before proceeding further, let me briefly summarise the derivation of this result, based on ref. \refcite{newper}.  
One can show that
when the metric changes by
$\delta g^{ab}=\nabla^a\xi^b+\nabla^b\xi^a$  the two pieces in the action  \eq{actfunc} changes by:
\begin{equation}
\delta A_{matt}
=-\int_\mathcal{V}d^4x\sqrt{-g}\nabla_a(T^a_b\xi^b); \quad \delta A_{\rm sur} = \frac{1}{8\pi  G} \int_{\mathcal{V}} d^4 x\,  \sqrt{-g}\, \nabla_a (R^a_b \xi^b)
\label{delmat}
\end{equation}
The first one arises from the definition of $T_{ab}$ when we use $\nabla_a T^a_b=0$
 which arises from the  equations of motion for the matter. The second one can be obtained from the fact that, the variation of the  surface term precisely cancels the variation $g^{ab}\delta R_{ab}$. (This idea, in fact, can be trivially generalised to any action which can be separated into bulk and surface terms; we will say more about this in Sec.\ref{sss:gb}.)

The integration of the divergences in Eq.(\ref{delmat})  over $\mathcal{V}$ leads to surface terms which contribute only on $\mathcal{H}$, since $\xi^a$ is nonzero only on the horizon.
Further, since $\xi^a$ is in the direction of the normal,
the demand $\delta A_{tot}=\delta A_{\rm sur} + \delta A_{\rm matter}=0$ leads to the result $(R^a_b-8\pi GT^a_b)\xi^b\xi_a=0$. Using the fact that $\xi^a$ is arbitrary \textit{except for being a null vector}, this requires  $R^a_b-8\pi GT^a_b=F(g)\delta^a_b$, where $F$ is an arbitrary function of the metric. Finally, since $\nabla_a T^a_b=0$ identically, $R^a_b-F(g)\delta^a_b$ must have  identical zero divergence; so $F$ must have the form
$F=(1/2)R+\Lambda$ where $R$ is the scalar curvature and $\Lambda$ is an undetermined integration constant.
The resulting equation is
\begin{equation}
R^a_b-(1/2)R\delta^a_b+\Lambda\delta^a_b=8\pi GT^a_b
\end{equation}
which is identical to Einstein's equation with an undetermined integration constant.  Since this should hold for the virtual displacements of every null surface $\mathcal{H}$ through every event (which will be a horizon  to some congruence $\mathcal{C}$), the field equations hold at every event.

This formalism ties up neatly with each of the issues raised in the last Section which are
worth stressing: 
\begin{itemize}
\item 
We now have a formalism in which gravity contributes on the boundary rather than in the bulk
(Sec. \ref{sss:dof}). It leads to the same equations as the bulk description essentially because of the holographic identity (Sec. \ref{sss:holo}). 
\item 
The surface term of
 gravitational action principle has a clear thermodynamic interpretation:
$A_{sur}$ is directly related to the (observer dependent horizon) entropy. For example, if we choose a local Rindler frame near the horizon with the Euclidean continuation for the metric:
\begin{equation}
ds^2_E\approx N^2 d\tau^2 +dN^2/\kappa^2+dL_\perp^2
\end{equation}  
then horizon maps to the origin and the region outside the horizon corresponds to $N>0$. 
This fits with our idea that  observers with a horizon should only use regions and variables accessible to them (Sec.\ref{sss:horizon}). The surface term can now be computed by integrating over the surface 
$N=\epsilon, 0<\tau<2\pi/\kappa$ and taking the limit  $\epsilon\to 0$. The unit normal is $u^a=\kappa(0,1,0,0)$ and $K=-\nabla_au^a$ so that:
\begin{equation}
A_{\rm sur}=-\frac{1}{8\pi G}\int d^2 x_\perp \int_0^{2\pi/\kappa}d\tau 
\epsilon \left( \frac{\kappa}{\epsilon}\right) =-\frac{1}{4}\frac{\mathcal{A}_\perp}{G}
\end{equation} 
where $\mathcal{A}_\perp $ is the area in the transverse directions.
Since the surface contribution is due to the existence of an inaccessible region we can identify $(-A_{\rm sur})$ with an entropy. 
\item 
The
variation of the surface term, when the horizon is moved infinitesimally, is equivalent to the change in the entropy $dS$ due to virtual work. The variation
 of the matter term contributes the $PdV$ and $dE$ terms and the entire variational principle is equivalent to the thermodynamic identity 
$TdS=dE+PdV$ applied to the changes when a horizon undergoes a virtual displacement. 
 This is exactly what we anticipated in Sec.(\ref{sss:tds}). 
 \item 
The semiclassical theory will depend on the wave functional $\exp iA_{sur}$, rather than on
$A_{sur}$. Foliation independence of this semiclassical limit is ensured if we demand that 
$
\exp iA_{sur} =\exp 2\pi i n
$.
This immediately leads to area quantization law:
$\mathcal{A}_\perp=(8\pi L_P^2) n
$. Such results have been around for some time now in different approaches to quantum gravity.
\item 
 Finally, the virtual displacements of horizon normal to itself, actually leads to the equation $(G_{ab}-8\pi T_{ab})\xi^a\xi^b=0$ for all \textit{null} vectors $\xi^a$. Again, this is what we needed to make the theory invariant under the transformation $T_{ab}\to T_{ab}+\Lambda g_{ab}$ to reformulate the cosmological constant problem (Sec. \ref{sss:gravimmu})
\end{itemize}

Since this approach has brought to center-stage the surface term, it is worth pointing out an important property of this term. In a general gauge, the $L_{sur}$ of the Einstein-Hilbert action has the form:
\begin{equation}
\sqrt{-g}L_{sur}=\frac{1}{2l_P}\partial_a[\sqrt{-g}(g^{bc}\Gamma^a_{bc}-g^{ab}\Gamma^c_{bc})]
\label{exactsur}
\end{equation} 
Consider now the linear approximation to Einstein gravity around flat spacetime
("graviton in flat spacetime") obtained by taking $g_{ab}=\eta_{ab}+l_P h_{ab}$ with $l_P^2=8\pi G$. (The dimension of $h_{ab}$ is $1/L$ as expected.) Then the Hilbert action has the structure:
\begin{equation}
 \sqrt{-g}L_{EH}=\frac{1}{2 l_P^2}\sqrt{-g}R\sim\frac{1}{2l_P^2}[(\partial g)^2+\partial^2g]
=\frac{1}{2}(\partial h)^2+\frac{1}{2l_P}\partial^2 h
\end{equation}
We see that the surface term is non-perturbative in $l_P$ in the ``graviton" picture! It follows that
staring from quadratic spin-2 action and iterating to all orders in the coupling constant $l_P$ can \textit{never} lead to a term which is non-analytic in $l_P$. So such an iteration can only lead to $L_{bulk}$ and \textit{not} to $L_{sur}$ or to $L_{EH}$. The claim, sometimes made in the literature, that 
 the Einstein-Hilbert action can be obtained  by starting from the action functional for spin-2 graviton, coupling it to its own stress tensor and iterating the process to all orders, is incorrect (for more details, see \refcite{gravitonmyth}).
Also note that the surface term at linear order
\begin{equation}
\sqrt{-g}L_{sur}\approx\frac{1}{2l_P}\partial_a\partial_b[\eta^{ab}h^i_i-h^{ab}]
\end{equation}
is invariant under the linear gauge transformations $h_{ab}\to h_{ab}+\partial_a\xi_b+\partial_b\xi_a$.  However, the exact form of $L_{sur}$ in \eq{exactsur}
is \textit{not} generally covariant. One might (incorrectly) think of general coordinate transformations as arising from the ``exponentiation" of infnitesimal co-ordinate transformations and it is sometimes (incorrectly) claimed 
that if a term is gauge invariant in the linear order, it will be generally covariant in the exact theory.  The $L_{sur}$ is a concrete counterexample. There is more to gravity than gravitons.

To understand the nature of the action principle based on the surface term, it is worth exploring whether such a formulation is possible for other theories, say, $U(1)$ gauge theory. The usual
action is:
\begin{equation}
L_{std}=-\frac{1}{16\pi} F_{ij}F^{ij}+ L_m
\end{equation} 
 Varying $A_j$ will give the Maxwell's equations $\partial_aF^{ab}=-4\pi J^b$ if we assume that
  $(F^{ab}\delta A_b n_a)=0$ on the boundary.
 Curiously enough, one can also get the same Maxwell's equations from the pure boundary action:
\begin{equation}
\mathcal{A}_{NEW}
=-\frac{1}{4\pi}\int_{\partial\mathcal{V}}  d^3x\ n_a(F^{ab}A_b)+\int_{\mathcal{V}}  d^4x L_m
\label{anew}
\end{equation} 
if we demand $\delta \mathcal{A}_{NEW}$ is zero for the gauge transformations $A_i\to A_i+\partial_i f$. The surface term in \eq{anew} is indeed the holographic dual of the bulk term (i.e., obtained by the relation in \eq{surbulkrel}) and the gauge transformations, under which we demand invariance, is analogous to the coordinate shifts $x^a\to x^a+\xi^a(x)$ used in the case of gravity. However, the similarity ends there. The surface term in \eq{anew} has no natural interpretation and is introduced rather arbitrarily. (This is because, unlike in gravity, the original action does not have second derivatives.) Needless to say, there is no thermodynamic analogue or blocking of information.

We shall now explore the last issue raised in Sec.(\ref{sss:guide}) to see why the formalism works and how it can be further generalised.

\section{Holographic structure of Semiclassical Action for Gravity}
\label{sss:gb}
There is actually a deep reason as to why this works, which  goes beyond the Einstein's theory. Similar results exists for \textit{a wide class of covariant theories based on principle of equivalence}, in which the gravity is described by
a metric tensor $g_{ab}$. Let me briefly describe the general setting from which this thermodynamic picture arises \cite{cc1}.

 Consider a  (generalized) theory of gravity in D-dimensions based on a generally covariant
scalar Lagrangian $L[g^{ab},R^a_{\phantom{a}bcd}]$ which is a functional of the metric $g^{ab}$ and curvature $R^a_{\phantom{a}bcd}$ and its covariant derivatives.
Instead of treating $[g^{ab},\partial_cg^{ab},
\partial_d\partial_cg^{ab}$] as the independent variables,  it is convenient to use $[g^{ab},
\Gamma^i_{kl},R^a_{\phantom{a}bcd}]$ as the independent variables. The curvature tensor $R^a_{\phantom{a}bcd}$ can be expressed entirely in terms of $\Gamma^i_{kl}$ and $\partial_j\Gamma^i_{kl}$ and is \textit{independent} of $g^{ab}$. 
To  investigate the general (``off-shell") structure of the theory, let us note that any scalar which depends on  $R^a_{\phantom{a}bcd}$ can be written in the form
$
L=Q_a^{\phantom{a}bcd}R^a_{\phantom{a}bcd}
$ with
the tensor $Q_a^{\phantom{a}bcd}$ depending on curvature and metric. (For example,   any function $L$ can be written trivially as 
$L=Q_a^{\phantom{a}bcd}R^a_{\phantom{a}bcd}$ with $Q_a^{\phantom{a}bcd}=(L/2R)(\delta^c_ag^{bd}-\delta^d_ag^{bc})$
but, of course, other choices of $Q_a^{\phantom{a}bcd}$ can also lead to the same $L$.)
Varying  the action functional will give, 
\begin{equation}
\delta A = \delta \int_\mathcal{V} d^Dx\sqrt{-g}\, L = \int_\mathcal{V} d^Dx \, \sqrt{-g} \, E_{ab} \delta g^{ab} + 
\int_\mathcal{V} d^Dx \, \sqrt{-g} \, \nabla_j \delta v^j
\label{deltaa}
\end{equation}
where $P_a^{\phantom{a}bcd}\equiv (\partial L/\partial R^a_{\phantom{a}bcd} )$ and
\begin{equation}
E_{ab}\equiv\left( \frac{\partial \sqrt{-g}L}{\partial g^{ab}} 
-2\sqrt{-g}\nabla^m\nabla^n P_{amnb} \right)
\label{eab}
\end{equation} 
and
\begin{equation}
\delta v^j \equiv [2P^{ibjd}(\nabla_b\delta g_{di})-2\delta g_{di}(\nabla_cP^{ijcd})]
\label{genvc}
\end{equation} 
This result is completely general. In $\delta A$ in \eq{deltaa}, the second term will lead to a surface contribution. 
To have a good variational principle leading to the result
$E^{ab} = $ matter source terms, we need to assume that $n_a \delta v^a =0$ on 
$\partial \mathcal{V}$ where $n_a$ is the normal to the boundary. In general this requires a particular combination of the ``coordinates" [$g_{ab}$] and the ``momenta" [$\nabla_c\delta g_{ab}$] to vanish and we need to put conditions on \textit{both} the dynamical variables \textit{and} their derivatives on the boundary.  
It is more reasonable (especially in a quantum theory) to choose either the variations of coordinates or those of momenta to vanish rather than a linear combination.  It is clear from \eq{deltaa} that this requires the condition $\nabla_a(\partial L/\partial R_a^{\phantom{a}bcd})=0$. When
$Q^{ijcd}$ is a polynomial in the curvature tensor (or, more generally, has a Taylor series expansion in the curvature tensor), we can instead use the condition: 
$
\nabla_cQ^{ijcd}=0
$
Because of the symmetries, this implies that $Q^{abcd}$ is divergence-free in \textit{all} indices. With this motivation, we shall hereafter confine our attention to Lagrangians of the form:
\begin{equation}
L=Q_a^{\phantom{a}bcd}R^a_{\phantom{a}bcd}; \qquad \nabla_cQ^{ijcd}=0
\end{equation} 
This may be thought of as a basic postulate on the form of the action.

To proceed further, it is useful to work with a
a set of tetrads $e^{\,c}_{(k)}$ where $k=0,1, ..., D$ identifies the vector and $c$
indicates the component. The dual basis is given by  $e_d^{(k)}$ with
$e_d^{(k)} e^{\,c}_{(k)} = \delta^c_d$.  Writing 
$R^{\,c}_{\phantom{\,c}dba} = e_d^{(k)} \nabla_{[b} \nabla_{a]}\,  e_{(k)}^{\,c}$ 
our Lagrangian becomes 
\begin{eqnarray}
L &=& Q_c^{\phantom{c}dba}\,  R^{\,c}_{\phantom{\,c}dba} = 2 Q_c^{\phantom{c}dba}\,  e_d^{(k)}\, \nabla_{b} \nabla_{a}\,  e^{\,c}_{(k)}\nonumber\\
&=& \nabla_b \left( 2 Q_c^{\phantom{c}dba}\,  e_d^{(k)}\, \nabla_a \, e^{\,c}_{(k)}\right)- 2 Q_c^{\phantom{c}dba}
\left( \nabla_b\,  e^{(k)}_d\right) \left( \nabla_a \, e^{\,c}_{(k)}\right)
\label{newr}
\end{eqnarray}
where we have done an integration by parts and used $\nabla_b Q_c^{\phantom{c}dba} =0$.
In a coordinate basis with 
$e^c_{(k)} = \delta^c_k,\nabla_a \, e^{\,c}_{(k)}=\Gamma^c_{ak},\nabla_b\,  e^{(k)}_d=-\Gamma^k_{bd}$ it reduces to the form:
\begin{equation}
\sqrt{-g}L=2\partial_c\left[\sqrt{-g}Q_a^{\phantom{a}bcd}\Gamma^a_{bd}\right]
+2\sqrt{-g}Q_a^{\phantom{a}bcd}\Gamma^a_{dk}\Gamma^k_{bc}\equiv L_{\rm sur} + L_{\rm bulk} 
\label{gensq}
 \end{equation} 
[More geometrically, writing
 $\mathcal{R}^a_{\phantom{a}b} = (1/2!) R^a_{\phantom{a}bcd} \, w^c \wedge \, w^d$ in terms of the  basis one forms $w^a$,
 introducing a corresponding two form for $Q_{abcd}$ with $\mathcal{Q}^a_{\phantom{a}b} =(1/2!) Q^a_{\phantom{a}bcd}\, w^c \wedge w^d$
 and using $\mathcal{R}^a_{\phantom{a}b} =  d\,   \Gamma^a_{\phantom{a}b} + \Gamma^a_{\phantom{a}c}\, \wedge \, \Gamma^c_{\phantom{c}b}$ where $\Gamma^a_{\phantom{a}b}$ are the curvature two forms.
 we can write the 
 Lagrangian as
 \begin{equation}
L = \ast \mathcal{Q}^a_{\phantom{a}b}\wedge \mathcal{R}^b_{\phantom{b}a} 
= \ast \mathcal{Q}^a_{\phantom{a}b}\wedge \left( d \Gamma^b_{\phantom{b}a} +  \Gamma^b_{\phantom{b}c}\wedge
\Gamma^c_{\phantom{c}a}\right) 
= d\left( \ast \mathcal{Q}^a_{\phantom{a}b}\wedge \Gamma^b_{\phantom{b}a}
\right) + \ast\mathcal{Q}^a_{\phantom{a}b}\wedge \Gamma^b_{\phantom{b}c}\wedge \Gamma^c_{\phantom{c}a}
\end{equation}
provided the $\mathcal{Q}^a_{\phantom{a}b}$ satisfies the condition:
$d\left( \ast \mathcal{Q}^a_{\phantom{a}b}\right) =0$ corresponding to $\nabla_c Q_a^{\phantom{a}bcd} =0$. The separation between bulk and surface terms is obvious. All these can, of course, be obtained by standard index gymnastics.]
 
 This result in \eq{gensq} shows that any such gravitational Lagrangian,
built from metric and curvature, has a separation into a surface term  and bulk term   in a natural but non covariant manner.  Ignoring the surface term, one can obtain the same \textit{covariant} equations of motion  $E_{ab} =T_{ab} $ even from a \textit{non covariant} bulk Lagrangian, $L_{bulk}$. What is more, there is
 a striking holographic relation between the surface and bulk terms in the Lagrangian
in \eq{newr}. It is easy to see that
\begin{equation}
L_{\rm sur} = - \frac{1}{2} \nabla_b \left[ e^{(k)}_d \frac{\partial L_{\rm bulk}}{\partial(\nabla_b e^{(k)}_d)}\right]
\end{equation} 
This is the same as the holographic identity in \eq{surbulkrel} in terms of the tetrads as the basic variables (see ref. \refcite{cc1}, section 5 for more details).
This remarkable result,  for a generic action functional in \eq{gensq}  makes all such  actions intrinsically holographic with the surface term containing an equivalent  amount of
information as the bulk. 
We mentioned earlier  (Sec. \ref{sss:holo}) that the bulk and surface terms of Einstein-Hilbert action
   are related
by this identity.
The current result shows that this is very general and is based only on the 
principle of equivalence (which allows the gravity to be described by a metric), general covariance
(which fixes the generic form of the action) and demand $\nabla_a Q^{abcd}=0$ (which is related to  the existence of a well defined  variational principle).

Everything else goes through as before  and it is possible
to reformulate the theory retaining \textit{only} the surface term for the gravity sector as in the
case of Einstein gravity.
The derivation given in ref. \refcite{newper} (especially Eqs. 7 to 9)  for Einstein-Hilbert action is easily generalisable for the general case  when $\nabla_a Q^{abcd}=0$. 
(For a related but alternative approach, see ref. \refcite{thomas}.)
 If one considers the infinitesimal virtual displacement $x^a\to x^a+\xi^a$ of the horizon, and use the fact that any scalar density changes by $\delta (\sqrt{-g}S)=-\sqrt{-g}\nabla_a(S\xi^a)$ one can obtain
 two key results: First is the identity 
$
\nabla_a E^{ab}=0
$
which is just the generalization of Bianchi identity. Second, if we consider an action principle with
based on $(A_m+A_s)$ where $A_m$ is the matter action and $A_s$ is the action obtained from $-L_{sur}$
(exactly as we did in the case of Einstein-Hilbert action; see discussion around \eq{actfunc}) then, for variations that arise from displacement of a horizon normal to itself, one gets the equation $(E_{ab}-\frac{1}{2}T_{ab})\xi^b\xi^a=0$ where $\xi^a$ is \textit{null}. Combined with  identity $\nabla_a E^{ab}=0$ this will lead to standard field equations with a cosmological term $E_{ab}=(1/2)T_{ab}+\Lambda g_{ab}$ just as in the case of Einstein-Hilbert action  \cite{newper}.

Since the basic equations are $(E_{ab}-\frac{1}{2}T_{ab})\xi^b\xi^a=0$, with $\xi^a$ being any null vector,
the addition of a cosmological constant
--- by the change $T_{ab}\to T_{ab}+\Lambda g_{ab}$ --- leaves the equations invariant. \textit{Gravity ignores the bulk vacuum energy density!}  When generalised Bianchi identity is used, once again the cosmological constant arises only as an integration constant;
 but then, it can be set to any value as a feature of the \textit{solution} to the
field equations in a given physical context. (It has a status similar to the  mass $M$ in the Schwarzschild metric).  This provides a basic reason for ignoring the bulk \cc\ 
\textit{and its changes} during various phase transitions in the universe.
When coupled to the  thermodynamic paradigm, which suggests  that
 in the presence of a horizon we should  work with the degrees of freedom confined by the horizon, it is possible to predict the value of this integration constant  \cite{cc1,tpcqglamda}.

Let us now consider the explicit form of divergence-free fourth rank tensor $Q_a^{\phantom{a}bcd}$, having the symmetries of the curvature tensor, 
 which determines the structure of the theory. The semiclassical,
low energy, action for gravity can now be determined from the derivative expansion
of $Q_a^{\phantom{a}bcd}$ in powers of number of derivatives: 
\begin{equation}
Q_a^{\phantom{a}bcd} (g,R) = \overset{(0)}{Q}_a{}^{bcd} (g) + \alpha\, \overset{(1)}{Q}_a{}^{bcd} (g,R) + \beta\, \overset{(2)}{Q}_a{}^{bcd} (g,R,\nabla R) + \cdots
\label{derexp}
\end{equation} 
where $\alpha, \beta, \cdots$ are coupling constants. We will treat the expansion in terms of the number of derivatives as giving the quantum corrections to the classical theory. To determine the first term, say, we only need to obtain all the possible fourth rank tensors $Q^{abcd}$ which (i) have the symmetries of curvature tensor; (b) are divergence-free and (iii) are made from $g^{ab}$; similarly, to obtain the next term, we allow the tensor $Q^{abcd}$ to depend on $g^{ab}$ and  $R^a_{\phantom{a}bcd}$ etc. 
Interestingly enough, at the first two orders, this leads to \textit{all} the gravitational theories (in D dimensions) in which
the field equations are no higher than second degree, though we did \textit{not }demand that explicitly. 

To see this, let us note that
if we do not use the curvature tensor, then we have just one unique choice for zeroth order,
 made from metric:
\begin{equation}
\overset{(0)}{Q}_a{}^{bcd}=\frac{1}{2}(\delta^c_ag^{bd}-\delta^d_ag^{bc})
\label{pforeh}
\end{equation} 
which satisfies our constraints.
When $Q_a^{\phantom{a}bcd}$ is built from metric alone,  \eq{gensq} becomes
\begin{equation}
\sqrt{-g}L=\partial_c\left[\sqrt{-g}(g^{bd}\Gamma_{bd}^c-g^{bc}\Gamma^a_{ba})\right]
+\sqrt{-g}(g^{bd}\Gamma^a_{dj}\Gamma^j_{ba}-g^{bc}\Gamma^a_{aj}\Gamma^j_{bc})
\label{surbulkeh}
\end{equation} 
which is precisely the bulk-surface decomposition for Einstein-Hilbert action. \footnote{The $Q^{abcd}$ for the Einstein-Hilbert case is closely related to the superpotential for energy momentum pseudotensor for gravity. In fact, in the locally inertial frame, one can re-express
$L=Q^{abcd}R_{abcd}\sim Q^{abcd}\partial_a\partial_c g_{bd}$ in terms of $L_{mod}\sim g_{bd}\partial_a\partial_cQ^{abcd}\sim g_{bd}t^{bd} $
where $t^{bd} $ is the gravitational energy momentum pseudotensor.
}
 
Next, if we allow for $Q_a^{\phantom{a}bcd}$ to depend linearly on curvature, then we have the following 
additional choice of  tensor with required symmetries:
\begin{equation}
\overset{(1)}{Q}_{abcd}=R_{abcd} -  G_{ac}g_{bd}+ G_{bc}g_{ad} + R_{ad}g_{bc} - R_{bd}g_{ac} 
\label{ping}
\end{equation} 
(In four dimensions, this tensor is essentially the double-dual of
$R_{abcd}$ and in any dimension can be obtained from $R_{abcd}$ using the alternating tensor  \cite{love}.)
In this case,  we get
\begin{eqnarray}
L&=&\frac{1}{2}\left(g_{ia}g^{bj}g^{ck}g^{dl}-4g_{ia}g^{bd}g^{ck}g^{jl}
+\delta^c_a\delta^k_ig^{bd}g^{jl}\right)R^i_{\phantom{i}jkl}R^a_{\phantom{a}bcd}\nonumber\\
&=&\frac{1}{2}\left[R^{abcd}R_{abcd}-4R^{ab}R_{ab}+R^2\right]
\end{eqnarray} 
This is just the Gauss-Bonnet(GB) action which is a pure divergence in 4 dimensions but not in higher dimensions.

The unified procedure for deriving Einstein-Hilbert action and GB action  shows that they are more closely related  to each other than previously suspected. The fact \cite{zw} that \textit{string theoretical models get GB type terms as corrections} is noteworthy in this regard. 
We can similarly determine the higher order corrections. More generally, one can construct the Lagrangian as a sum of terms, each 
of which is a homogeneous function of degree $m$ which are divergent-free. This leads one to the  Lovelock Lagrangian \cite{love}.
Also note that, as long as the  higher order quantum gravitational corrections are determined by the holographic principles, the higher orders terms will all respect the invariance of the theory under $T_{ab}\to T_{ab}+\lambda g_{ab}$ and the cosmological constant will continue to remain an integration constant even when quantum corrections are incorporated.

 \section{Conclusions}
 
I have taken a rather deductive approach in the above discussion, raising a set of issues related to conventional approach to gravity and
showing how they lead to a new perspective based on surface degrees of freedom. The natural description incorporating this perspective leads to results which go beyond the conventional Einstein gravity, and provides a handle on the  semiclassical corrections. In this last section, it is useful to take a complementary inductive path and (necessarily) speculate on the broader picture.

Such a broader picture is clearly the one in which the
 continuum spacetime is like an elastic solid (`Sakharov paradigm'; see e.g. ref. \refcite{sakharov}) with Einstein's equations providing the macroscopic description. 
 In this approach, the full theory has some microscopic variables $q_i$ and an  action $A_{micro}(q_i)$. Integrating out short wavelength fluctuations and microscopic degrees of freedom  should lead to a  long wavelength effective action, which should be a pure surface term, obtained in \eq{gensq}:
 \begin{equation}
\sum\exp(-A_{micro})\propto \exp(-A_{sur})
\propto \exp\left(-2\int_{\partial\mathcal{V}} d^{D-1}x\sqrt{-g}\ n_cQ_a^{\phantom{a}bcd}\Gamma^a_{bd}\right)
\label{eact}
\end{equation} 
 as well as bring about $g_{ab}$ as the new dynamical variables in terms of which the effective action is described. 
 This suggests that \textit{the effective low energy  degrees of freedom of gravity
for a volume $\mathcal{V}$
reside in its boundary $\partial\mathcal{V}$} --- a  point of view that is strongly supported by the study
of horizon entropy, which shows that the degrees of freedom hidden by a horizon scales as the area and not as the volume.

 In this description $g_{ab}$ is like the density $\rho$ of a solid arising from large number of atoms and is not a fundamental dynamical variable. It does not make sense to vary the
 $g_{ab}$ in this action.
 Instead, the (covariant) equations of motion are obtained by demanding the 
invariance of the (noncovariant) surface action $A_{sur}$ in \eq{eact},
 under virtual displacements of any (observer dependent) horizon normal to itself.
 This might seem unusual at first but, as I explained in Sec.\ref{sss:tds}, it arises from the thermodynamic interpretation of (observer dependent) horizons. The discussion
 in Sections \ref{sss:foliation} to \ref{sss:tds} as well as in Sec.\ref{sss:holo} clearly points to such a ``surface-based" approach
and all the issues raised there get a natural interpretation in this approach. In the displacement $x^a\to \bar{x}^a=x^a+\xi^a$ the
$\xi^a(x)$ is similar to the displacement vector field used, for example,
in the study of elastic solids. 
The true degrees of freedom are some unknown `atoms of spacetime' but in the continuum limit,
the displacement $x^a\to \bar{x}^a=x^a+\xi^a(x)$ captures the relevant dynamics,  just like 
in the study of elastic properties of the continuum solid. In fact, one can reformulate the Einstein gravity in terms of
the dynamics of a null vector field in a background spacetime \cite{elastic}.
The horizons in the spacetime are then  similar to defects in the solid so that their displacement costs entropy. 
\footnote{This is easily seen from \eq{newr} in which   $L_{bulk}$ can be thought of as describing the dynamics of four vector fields  $e^{(k)}_d$
with $k=0,1,2,3$. If one of them, say  $k=0$, is a null vector $\xi^d\equiv e_{(0)}^d$ then the Lagrangian for this vector, on using
 \eq{pforeh}, becomes $L_{bulk}\propto [\nabla_a\xi^a\nabla_b\xi^b -\nabla_a\xi^b\nabla_b\xi^a$].
 This is what I used in ref. \cite{elastic}, without the null condition, to translate the dynamics of Einstein gravity to the dynamics of a  vector displacement field. The collective dynamics of all null surfaces leads to the standard Einstein gravity.}

\textit{The approach also leads to two new insights, which one could not have been anticipated a priori.} 

(1) The surface action leads in a natural fashion to equations
$(E_{ab}-T_{ab})\xi^a\xi^b=0$ for all null vectors $\xi^a.$  These equations of motion are now invariant
under the changes to the vacuum energy $T_{ab}\to T_{ab}+\Lambda g_{ab}$ and we have a natural solution to the \cc\ problem.
Note that, this approach, unlike many others, can handle the \textit{changes} to the vacuum energy density arising due to phase transitions in the early universe. The observed \cc\ can be interpreted \cite{cc2}  as arising due to the vacuum fluctuations in a region confined by the horizon and --- in that sense --- is coupled to the surface degrees of freedom of gravity. 

(2) The effective action in \eq{eact} can be expanded in terms of number of derivatives and the
low energy effective action for gravity is then determined by the derivative expansion
of $Q_a^{\phantom{a}bcd}$ in powers of number of derivatives, given by \eq{derexp}:
\begin{equation}
e^{-A_{sur}}
\propto \exp\left(-2\int d^{D-1}x\sqrt{-g}\ n_c
[
\overset{(0)}{Q}_a{}^{bcd} (g) + \alpha\, \overset{(1)}{Q}_a{}^{bcd} (g,R) + \cdots
]
\Gamma^a_{bd}\right)
\end{equation} 
The first term is the surface term corresponding Einstein-Hilbert action and the second one leads to the Gauss-Bonnet action; this (as well as higher order terms) have a natural interpretation of being a quantum correction in this approach. 
We also have a general principle for determining
the correction terms (by constructing the divergence free tensor $Q_a^{\phantom{a}bcd}$ from variables with right number of derivatives) and constraining the structure of underlying theory. 
It is worth recalling \textit{that such a Gauss-Bonnet term
arises as the correction in string theories} \cite{zw}. The thermodynamic interpretation (which is on-shell) as well as
the holographic description (which is off-shell) are also applicable to quantum corrections
to the Einstein-Hilbert Lagrangian. 
The invariance of the theory under $T_{ab}\to T_{ab}+\Lambda g_{ab}$ continues to hold for the higher order terms as well suggesting that the mechanism for ignoring the bulk \cc\ is likely to survive quantum gravitational corrections.

\section*{Acknowledgment}
I thank K.Subramanian for a careful reading for the manuscript.


\end{document}